\documentclass[epj]{svjour}
\usepackage{graphics}
\usepackage{threeparttable}
\begin{document}
\title{Linearly polarised photon beams at ELSA}
\subtitle{and measurement of the beam asymmetry in {\boldmath$\pi^0$}-photoproduction off the proton}
\author{\mail{D. Elsner, Nussallee 12, 53115 Bonn, Germany,
        \email{elsner@physik.uni-bonn.de}}
          D. Elsner\inst{1}, 
        B. Bantes\inst{1},
        O. Bartholomy\inst{2},
        D.E. Bayadilov\inst{2,3},
        R. Beck\inst{2},
        Y.A. Beloglazov\inst{3},
        R. Castelijns\inst{5}\fnmsep\thanks{present address: FZ J\"ulich, Germany},
        V. Crede\inst{2,6},
        A. Ehmanns\inst{2},
        K. Essig\inst{2},
        R. Ewald\inst{1},
        I. Fabry\inst{2},
        K. Fornet-Ponse\inst{1},
        M. Fuchs\inst{2},
        C. Funke\inst{2},
        A.B. Gridnev\inst{3},
        E. Gutz\inst{2},
        S. H\"offgen\inst{1},
        P. Hoffmeister\inst{2},
        I. Horn\inst{2},
        I. Jaegle\inst{8},
        J. Junkersfeld\inst{2},
        H. Kalinowsky\inst{2},
        Frank Klein\inst{1},
        Friedrich Klein\inst{1},
        E. Klempt\inst{2},
        M. Konrad\inst{1},
        M. Kotulla\inst{8,9},
        B. Krusche\inst{8},
        H. L\"ohner\inst{5}, 
        I.V. Lopatin\inst{3},
        J. Lotz\inst{2},
        S. Lugert\inst{9},
        D. Menze\inst{1},
        T. Mertens\inst{8},
        J.G. Messchendorp\inst{5},
        V. Metag\inst{9},
        C. Morales\inst{1},
        M. Nanova\inst{9},
        D.V. Novinski\inst{2,3},
        R. Novotny\inst{9},
        M. Ostrick\inst{1}\fnmsep\thanks{present address: University of Mainz,
        Germany},
        L.M. Pant\inst{9}\fnmsep\thanks{on leave from Nucl. Phys.
        Division, BARC, Mumbai, India},
        H. van Pee\inst{2},
        M. Pfeiffer\inst{9},
        A.V. Sarantsev\inst{2,3},
        C. Schmidt\inst{2},
        H. Schmieden\inst{1},
        B. Schoch\inst{1},
        S. Shende\inst{5},
        A. S{\"u}le\inst{1},
        V.V. Sumachev\inst{3},
        T. Szczepanek\inst{2},
        U. Thoma\inst{2},
        D. Trnka\inst{9},
        D. Walther\inst{1},
        C. Weinheimer\inst{2}\fnmsep\thanks{present address: University of
        M\"unster, Germany},
        \and C. Wendel\inst{2}
\newline(The CBELSA/TAPS collaboration)
}
\institute{Physikalisches Institut der Universit\"at Bonn, Germany 
           \and Helmholtz-Institut f\"ur Strahlen- und Kernphysik 
                der Universit\"at Bonn, Germany
           \and Petersburg Nuclear Physics Institute, Gatchina, Russia
           \and Physikalisches Institut, Universit\"at Erlangen, Germany
           \and KVI, University of Groningen, The Netherlands
           \and Department of Physics, Florida State University, Tallahassee,
           USA
           \and Physikalisches Institut, Universit\"at Basel, Switzerland
           \and II. Physikalisches Institut, Universit\"at Giessen, Germany
          }
%
\date{Received: date / Revised version: date}
%
\abstract{
At the electron accelerator ELSA a linearly polarised tagged photon beam is
produced by coherent bremsstrahlung off a diamond crystal.
Orientation and energy range of the linear polarisation can
be deliberately chosen by accurate positioning of the crystal with a
goniometer.
The degree of polarisation is determined by the form of the scattered electron spectrum. 
Good agreement between experiment and expectations on basis of the 
experimental conditions is obtained. Polarisation degrees of P$_{\gamma}$=
40\% are typically achieved at half of the primary electron energy. 
The determination of P$_{\gamma}$ is confirmed by measuring the beam asymmetry, $\Sigma$, in
$\pi^0$ photoproduction and a comparison of the results to independent measurements using laser
backscattering.
\PACS{
  {13.60.-r}{Photon and charged-lepton interactions with hadrons} \and
  {13.60.Le}{Meson production} \and
  {13.88.+e}{Polarization in interactions and scattering}
  } 
} 
\authorrunning{D. Elsner {\it et al.}}
\titlerunning{Linearly polarised photon beams at ELSA}
\maketitle
\section{Introduction}
\label{intro}
Experiments based on photo-induced exclusive reactions 
are well suited to improve our understanding of the complicated structure of
the nucleon. In addition to measurements of cross sections, polarisation observables
are indispensable. They are sensitive to interference terms and therefore give
access to small amplitudes, even if those are too small to affect the total cross
section significantly. Circularly and linearly polarised photon beams
allow, in combination with target or recoil polarisation, the extraction of double polarisation
observables. From linearly polarised photons alone the beam asymmetry,
$\Sigma$, can be extracted (for a definition of the observables see {\it e.g.} ref. \cite{Knoechlein95}).
In case of pseudoscalar meson photoproduction the beam asymmetry alone 
does not allow an unambiguous extraction of all partial waves \cite{Elsner07}, but
its measurement is essential in view of a complete experiment \cite{CT97}.\\
The two common methods for generation of  linearly polarised photons are
coherent bremsstrahlung and Compton backscattering (CBS).
In Compton backscattering the electron beam collides with a laser beam of
short wavelength. Linearly polarised photons can be produced using linearly polarised laser
photons \cite{Nakano01,GraalPi}.
The degree of polarisation of the CBS photons is proportional to that of laser
photons, with its maximum at the highest photon energy. At the present
facilities this method is compared to electron bremsstrahlung limited in intensity and achievable maximum photon
energy.\\
In coherent electron bremsstrahlung the recoil is transferred to a crystal
radiator. Depending on its orientation relative to the electron beam, the whole crystal
absorbs the recoil, which fixes the plane of electron
deflection. Consequently, the photons produced by the coherent process are
linearly polarised. Compared to CBS, photon beams from coherent bremsstrahlung have a higher
intensity, but on the other hand a lower maximum degree of polarisation at
higher photon energies.\\
Several facilities successfully use coherent bremsstrahlung to produce
linearly polarised photons at high energies \cite{Lohmann94,Cole02}.  
For the first time a setup for coherent bremsstrahlung was installed and
operated at the electron accelerator ELSA \cite{Elsa06}.
The following section is first devoted to the basics of coherent
bremsstrahlung. The description of the apparatus is then followed by the
alignment procedure for the crystal and the results obtained for the photon polarisation.
In sect. 6 the measurement of the photon beam-asymmetry, $\Sigma$, in
$\pi^0$-photoproduction is presented as an independent cross check for the
polarisation analysis.

\section{Coherent Bremsstrahlung}
\label{sec:CohBrems}
Radiators with a periodical lattice structure 
allow the  production of
linearly polarised photons via the process of coherent bremsstrahlung. 
In this section properties of the coherent process
are described which are essential 
for the understanding of the experimental
methods. For a more detailed discussion we refer to review
articles \cite{Ueberall55,Ueberall56,Ueberall57,Ueberall62,Diambrini68,Timm69}.\\
In the case of incoherent bremsstrahlung (bs), an electron with energy $E_0$
and momentum $\vec{p_0}$ radiates a photon with energy $k$, due to coulomb
interaction. Momentum conservation requires a
recoil partner to take over the recoil momentum
\begin{eqnarray}
\vec{q} = \vec{p_0} - \vec{p} - \vec{k}.
\end{eqnarray}
Here $\vec{p}$ denotes the momentum of the outgoing electron.
In general, a kinematical constraint applies for the longitudinal, $q_l$, and
transversal, $q_t$,  momentum transfer. A good approximation for this so-called
"pancake" condition \cite{Diambrini68,Timm69} is given by the relations
\begin{eqnarray}
 \delta \leq q_l \leq 2\delta \ \\
  0 \leq q_t \leq 2x . 
\label{eq:pancake_ql}
\end{eqnarray}
The longitudinal momentum transfer shows a non-zero lower limit given by
\begin{eqnarray}
  \delta(x) \equiv : q_l^{min} = \frac{1}{2E_0}\frac{x}{1-x},
   \label{eq:delta}
\end{eqnarray}
with the fractional photon energy x = k/$E_0$. 
\begin{figure}
  \begin{center}
    \resizebox{1.0\columnwidth}{!}{%
      \includegraphics{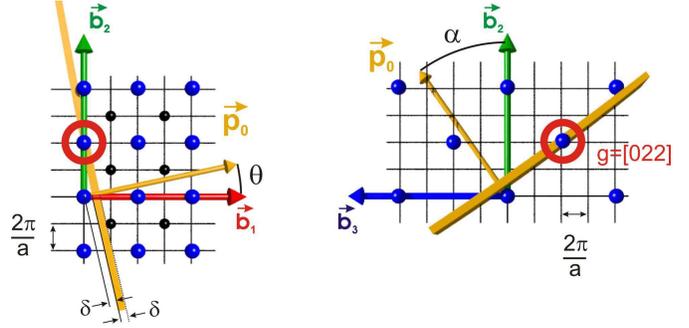}
    }
  \end{center}
  \caption{Left: Projection of four parallel planes through the reciprocal
    lattice vectors [000], [001], [002] and [003] in the plane
    $\vec{b_1}$, $\vec{b_2}$. The primary momentum $\vec{p_0}$ is rotated by
    a small angle $\theta$. The pancake region is indicated by a grey
    band. With regard to the $\vec{b_3}$-axis, perpendicular to the drawing
    plane, many vectors still lie in the allowed kinematical region
    (circle). Right: Reciprocal lattice vectors in the plane $\vec{b_2}$,
    $\vec{b_3}$. Compared to the left picture the angle $\theta$ is larger and
    $\vec{p_0}$ is tilted out of the plane $\vec{b_1}$, $\vec{b_2}$ by the
    angle $\alpha$, such that only the vector [02$\overline{2}$] lies in
    kinematical allowed region.
  }
  \label{fig:pancake}       
\end{figure}

In the case of incoherent bs only one single nucleus (or electron)
absorbs the momentum transfer, in contrast to the coherent process 
where the whole lattice participates, comparable to the M\"o\ss{}bauer effect.  
The process of coherent bs depends decisively on the orientation of the momentum transfer \vec{q} in
the reciprocal lattice space, more precisely, the momentum transfer has to fit
to a vector of the reciprocal lattice.
Consequently, only discrete recoil momenta can be transferred to the crystal
as a whole, specified by the Laue condition 
\begin{eqnarray}
\vec{q} = n \cdot \vec{g}. 
\end{eqnarray}
The reciprocal lattice vector
\begin{eqnarray}
\vec{g} = \sum \limits_{k=1}^3 h_k \vec{b_k}
\end{eqnarray}
is composed of the Miller indices, $h_k$, and the basis vectors of the
reciprocal lattice, $b_k$ (below we use the nomenclature [h$_1$h$_2$h$_3$]). Figure \ref{fig:pancake} shows the momentum vector
$\vec{p_0}$ in the reciprocal lattice space. For the selection of only one
reciprocal lattice vector in the kinematical region of allowed recoils the
angles $\theta$ in the plane $\vec{b_1b_2}$ and $\alpha$ in the plane
$\vec{b_2b_3}$ have to be carefully chosen. The pancake region is illustrated by
the grey band perpendicular to the momentum vector $\vec{p_0}$.

Contributions to the coherent bs cross section only result from reciprocal lattice vectors
within the pancake region. At a fixed orientation of the lattice, the pancake
shifts with increasing photon energy, cf. eq. \ref{eq:delta}.
Consequently, at a certain point a reciprocal lattice vector leaves the pancake.
This leads to a discontinuity in the coherent bs intensity at the
fractional photon energy 
\begin{eqnarray}
x_d = \frac{2 E_0 \delta}{(1 + 2 E_0 \delta)}.
\end{eqnarray}
The plane of the electron deflection is fixed very tightly by the incoming
electron momentum $\vec{p_0}$ and the reciprocal lattice vector $\vec{g}$ responsible
for the coherent process. Hence, the linear polarisation of the emitted
photons is oriented in the plane ($\vec{g},\vec{p_0}$). 

\section{Apparatus}
\label{sec:setup}
\begin{figure}
  \begin{center}
    \resizebox{1.0\columnwidth}{!}{%
      \includegraphics{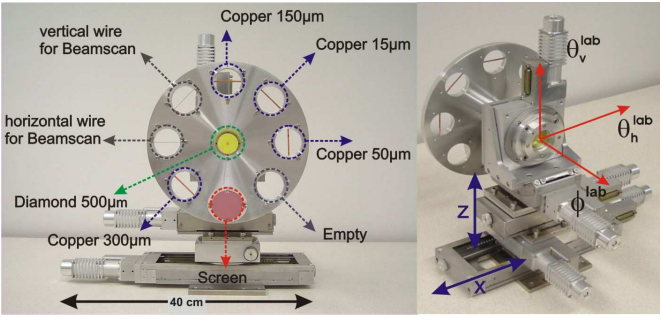}
    }
  \end{center}
  \caption{Goniometer setup for the Crystal-Barrel/TAPS experiment at
    ELSA. Left: Available  amorphous radiators which can be selected by rotation
    of the azimuthal axis. The horizontal translation allows the
    choice of either crystal or amorphous radiator. Right: Three step-motor
    drives may be used to rotate the crystal around a vertical axis
    ($\theta_v^{lab}$), a horizontal axis ($\theta_h^{lab}$) and an azimuthal
    axis ($\phi^{lab}$). The diamond crystal is mounted in the center of the
    three axes with its [100] crystal axis parallel to the goniometer
    axis ($\phi^{lab}$). The remaining two axes allow the horizontal and
    vertical translation.
}
  \label{fig:goniometer}       
\end{figure}

\begin{figure}[b]
  \begin{center}
    \resizebox{0.9\columnwidth}{!}{%
      \includegraphics{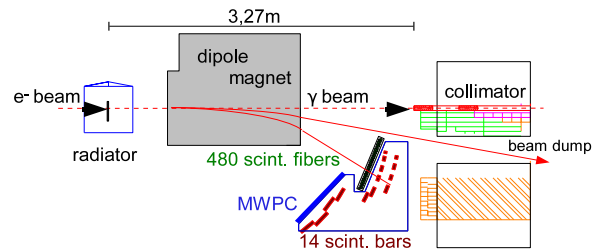}
    }
  \end{center}
  \caption{Setup of the tagging-system as described in the text. 
 }
  \label{fig:tagger}       
\end{figure}

The electron beam of ELSA \cite{Elsa06} hits the radiator target in front of
the tagging system \cite{KFP}. Electron beams of $E_0$ = 2.4\,GeV and
$E_0$ = 3.2\,GeV were routinely used for experiments with linearly polarised photons.\\
The crystal has a front surface of (4 x 4)\,mm and thickness of $500\,\mu$m.  
It is accurately positioned by a dedicated commercial 5-axis
goniometer\footnote{Newport Corporation, Irvine, CA 92606, USA}.
The maximum angular uncertainty is $\delta < 170\,\mu$rad due to the wobble
along the axes. Optical test measurements showed that other uncertainties,
like absolute accuracy, uni-directional repeatability and reversal value
(hysteresis) are negligible. 
The crystal, glued on a $12.5\,\mu$m kapton foil, is positioned in the common center
of the three rotation axes of the goniometer, horizontal ($\theta_h^{lab}$), vertical
($\theta_v^{lab}$) and azimuthal ($\phi^{lab}$), cf. fig. \ref{fig:goniometer}.
The minimum incremental motion (the smallest increment of motion the device is
capable of consistently and reliably delivering) of each rotation axes is
$\theta$ = 0.001\,degree. Special measurements, discussed in sect.
\ref{sec:stone}, confirm the orientation of the [100] crystal axis perpendicular to its front
surface.\\
Further copper radiators with different thicknesses and wires to scan
size and position of the electron beam are mounted
on a disk around the crystal, as can be seen in fig. \ref{fig:goniometer}.\\
Electrons which radiated a photon are momentum analysed using the
tagging-spectrometer, as schematically depicted in fig. \ref{fig:tagger}. 
The detection system consists of 14 plastic scintillators providing fast
timing and additional hodoscopes, a 480 channel scintillating fibre detector and
a MWPC, to achieve a good energy resolution.
The optimisation and analysis of linear polarisation is solely 
based on the data of the scintillating fibre detector which
covers the energy range $E_\gamma = 0.18 ... 0.8\,E_0$. 
The fibres are arranged in two layers. The individual fibres overlap by around
2/3 of their diameter. The energy resolution varies between 2\,MeV for the high photon energies
and 25\,MeV for the low energies for $E_0 = 3.2$\,GeV.
The tagged photon beam remains virtually uncollimated.
Hence, the measured electron spectrum directly reflects the photon spectrum.
The orientation of the linear polarisation and the 
position of the coherent maximum in the photon energy-spectrum depends on the
alignment of the crystal relative to the electron beam
direction. Maximum polarisation is found in the plane ($\vec{g},\vec{p_0}$). 
The direction of the momentum vector $\vec{p_0}$ in the reference frame
of the crystal is defined by the polar angle, $\theta$, and the azimuthal
angle, $\alpha$, cf. fig. \ref{fig:pancake}. The angle $\phi$ is the azimuth
of the reciprocal lattice vector $\vec{g}$ in the same reference frame.
Given $\theta$ and $\alpha$, the position of the discontinuity, $x_d$, in the energy-spectrum can be
calculated \cite{Timm69,Lohmann94}:  
\begin{eqnarray}
x_d = [2E_0(g_1\cos\theta + \sin\theta(g_2\cos\alpha + g_3\sin\alpha))]
\quad \quad \quad \\ \nonumber  
\cdot [1 + 2E_0(g_1\cos\theta + \sin\theta(g_2\cos\alpha + g_3\sin\alpha))]^{-1}.
\label{eq:disc-angle}
\end{eqnarray}
Relations between crystal angles ($\theta, \alpha, \phi$) and goniometer angles
($\theta_h^{lab}$, $\theta_v^{lab}$, $\phi^{lab}$) for the g=[02$\overline{2}$]
reciprocal lattice vector are given by \cite{Lohmann94}
\begin{eqnarray}  
\theta_v^{lab} & = & \arcsin(\sin\theta\ \sin(\alpha + \phi^{lab}));\\
\theta_h^{lab} & = & -\arctan(\tan\theta\ \cos(\alpha + \phi^{lab}));\\
\theta & = & \arccos(\cos\theta_h^{lab}\ \cos\theta_v^{lab});\\
\alpha & = &\arccos[(-\cos\phi^{lab} \sin\theta_h^{lab}
\cos\theta_v^{lab}\nonumber \\
& + & \sin\phi^{lab}\sin\theta_v^{lab})(\sin\theta)^{-1}];\\
\phi & = &\phi^{lab} + \beta.
\label{eq:calc_angles}
\end{eqnarray}
The angle $\beta$ defines the orientation of the polarisation plane, $\beta =
0$ and $\pi/2$ are associated to vertical and horizontal linear polarisation, respectively. 
It is essential to determine all angular offsets between the crystal reference
frame and the goniometer system on the one hand, and the incoming electron
beam and the goniometer system on the other hand with sufficient accuracy.
The former offsets have to be
measured once in case of a fixed installation. The latter depend on
the stability of the beam alignment and have to be determined repeatedly.
For the alignment process we use the coherent bremsstrahlung itself as
described in the next section.

\section{Crystal alignment - Stonehenge Technique}
\label{sec:stone}
\begin{figure}
  \begin{center}
    \resizebox{0.70\columnwidth}{!}{%
      \includegraphics{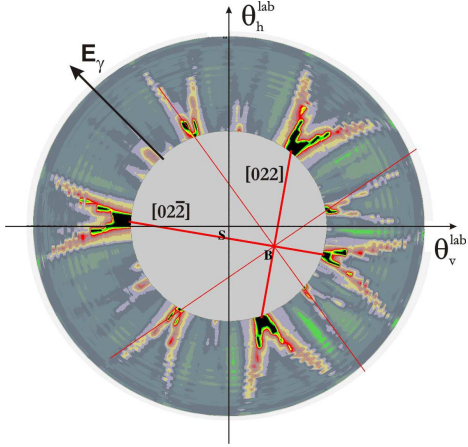}
    }
  \end{center}
  \caption{Stonehenge plot for non aligned crystal at an electron-beam energy of E$_0$=3.2\,GeV. The
    tagged photon energy is plotted radially, the colour-code (greyscales) represents the
    coherent intensity. The axes show the horizontal and vertical rotation in
    the goniometer system. The opening cone of the quasi azimuthal scan is $\theta_c=60\,$mrad, see
    eq. \ref{eq:quazi_azi}. The prominent structures associated to the [022] and
    [02$\overline{2}$] reciprocal lattice vectors. The structures rotated by 45
    degrees result from the [044] and [04$\overline{4}$]
    vectors and give additional interpolation points .  
}
  \label{fig:stone_off}       
\end{figure}

The alignment is achieved by the {\it Stonehenge Technique}. The procedure can cope with a relatively large mounting
misalignment and allows any arbitrary orientation of the polarisation plane to be selected.
A detailed description is given in \cite{StoneKen,StoneKen2}.
The basis of the technique is a quasi azimuthal scan which sweeps the crystal
axes in a cone of angular radius $\theta_c$ by stepping simultaneously on the
horizontal and vertical axis of rotation:
\begin{eqnarray}
\theta_v^{lab} & = & \theta_c\cos\Phi;\quad  0\leq\Phi < 2\pi \\
\theta_h^{lab} & = & \theta_c\sin\Phi.
\label{eq:quazi_azi}
\end{eqnarray}
For each point of the scan a photon energy spectrum is measured and plotted in
a polar diagram ({\it Stonehenge plot}), where the photon energy increases in
the outward radial direction and the x- and y-directions correspond to the
rotation axes $\theta_v^{lab}$ and $\theta_h^{lab}$, respectively.
In order to accentuate the coherence effect, the spectra are normalised to the
spectrum of an amorphous copper radiator. Therefore the colour code of the diagram
denotes the coherent intensity.
A Stonehenge plot for a non aligned crystal taken for the Crystal-Barrel/TAPS experiment
\cite{Elsner07} is shown in fig. \ref{fig:stone_off}.
The coherent contributions from different settings of crystal planes result in
pronounced structures due to the different angles between crystal and electron beam.
The strongest intensities typically relate to the [022] and [02$\overline{2}$] reciprocal
lattice vectors and the points where they converge at E$_{\gamma}\rightarrow$ 0 (inner
circle) indicate where the corresponding setting of planes is parallel to the
electron beam. The analysis of the symmetry in the Stonehenge plots yields
all independent offsets of the crystal. For a detailed description of this
analysis method it is referred to \cite{StoneKen,StoneKen2}. \\
Taking into account the angular offsets, it is
possible to set the linear polarisation at any desired spatial direction and
at any photon energy by choosing the crystal orientation.
\begin{figure}[t]
  \begin{center}
    \resizebox{1.0\columnwidth}{!}{%
      \includegraphics{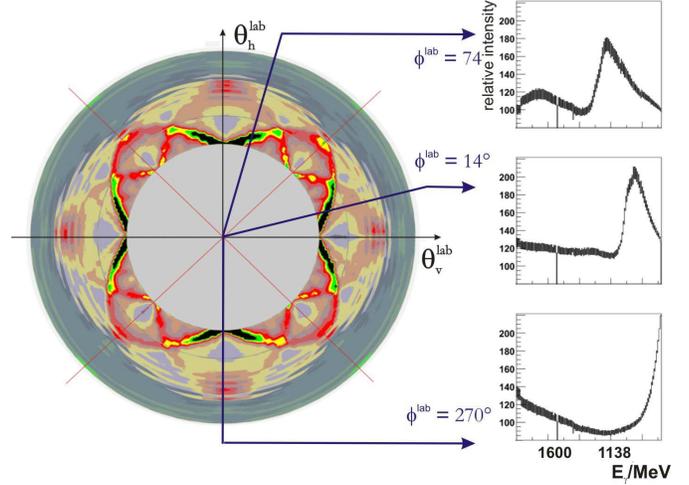}
    }
  \end{center}
  \caption{Stonehenge plot for perfect aligned crystal at E$_0$=3.2\,GeV,
    cf. fig. \ref{fig:stone_off}. Here the opening cone of the quasi azimuthal scan
    is $\theta_c=10\,$mrad. Photon energy spectra for three different
    $\phi$-angles are plotted additionally to illustrate the shifting of the
    coherent peak. }
  \label{fig:stone_ok}       
\end{figure}
Figure \ref{fig:stone_ok} shows the Stonehenge plot of a perfectly aligned
crystal with a vertical orientation of the [022] plane. Three  
one-dimensional sample histograms for different directions are shown on the right side.
The precision of the angular offsets depends on scan parameters, basically the
step size and the cone $\theta_c$ (cf. eq. \ref{eq:quazi_azi}). The resulting
azimuthal orientation has an accuracy of $\Delta\phi$ = 0.5 degree \cite{StoneKen2}.\\
In order to preserve the alignment during the experiment, the stability of the beam position 
is monitored online, using the coherent peak itself, since the position of the
coherent peak in the energy spectrum is extremely sensitive to angle of the
incident beam.
As for the Stonehenge plots, also for beam diagnostics the
coherent spectrum is always normalised to the spectrum of an amorphous
copper radiator. 
At the beginning of each experiment a normalised reference
histogram is defined, which is compared with the online spectrum permanently.
The incoherent copper spectrum is measured in regular intervals. 

\section{Degree of linear polarisation}
\label{sec:data}
\begin{figure}
  \begin{center}
    \resizebox{0.8\columnwidth}{!}{%
      \includegraphics{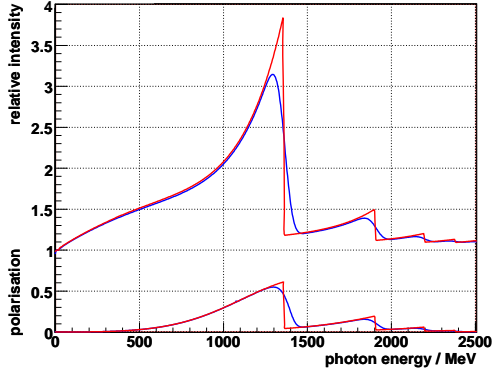}
    }
  \end{center}
  \caption{Comparison of the polarisation contribution of the [04$\overline{4}$] vector,
    with its maximum at E$_\gamma$=1900MeV, underneath the selected
    [02$\overline{2}$] vector at E$_\gamma$=1300MeV. The curve with the broadened peaks shows
    the effect of typical experimental conditions, basically multiple scattering
    and electron-beam divergence.}
  \label{fig:anb_coh}       
\end{figure}

The generation of high degrees of linear polarisation 
requires the isolated contribution of one of the [0,$\pm$2,$\pm$2,] reciprocal lattice
vectors to coherent bremsstrahlung. 
Precisely determined offsets (cf. sect. 4) enable to deliberately set both,
the energy of the coherent peak and the orientation of the linear polarisation.
The determination of the polarisation degree is based on the comparison of the
measured electron spectrum with the ANB (``analytic bremsstrahlung
calculation'') software \cite{ANB} from T\"ubingen University. The ANB code allows the calculation  
of coherent intensities for each single reciprocal lattice vector. It integrates over all desired vectors.
Taking into account the incoherent contributions, the degree of polarisation
can be determined. 
If there is no overlap of different reciprocal lattice vectors within a given
energy interval, the degree of polarisation can be obtained from {\it any} fit
of the intensity spectrum.\\
Figure\,\ref{fig:anb_coh} shows the ANB-calculated relative
photon intensity spectrum along with the calculated photon polarisation.
Due to a tiny overlap with the adjacent peak of the [04$\overline{4}$] vector the exact
determination of the polarisation degree from the experimental data alone is
not feasible. In our particular case no clear separation of the vector
[02$\overline{2}$] was realisable in the required energy region.

Under given experimental conditions the shape of the coherent spectrum is not
determined by the crystal orientation alone. Each single process has a small deviation
from the nominal kinematics as a result of beam divergence and multiple
scattering in the crystal. Both effects cause a smearing of the sharp edge at
the discontinuity position, due to the different
pancake conditions for individual processes, and hence lower the intensity and
consequently the maximum degree of polarisation (cf. fig. \ref{fig:anb_coh}).
The effects of beam divergence, beam spot-size, energy resolution and multiple scattering are
included in the ANB software. 
Tables \ref{tab_beam}-\ref{tab_goni_angles} show overviews of typical values
for our experimental parameters. The electron-beam energy and the spot size
are precisely measured. The values of the beam divergence result from
calculations of the beam-line optic. 
\begin{figure*}
\begin{center}
    \resizebox{0.95\textwidth}{!}{%
      \includegraphics{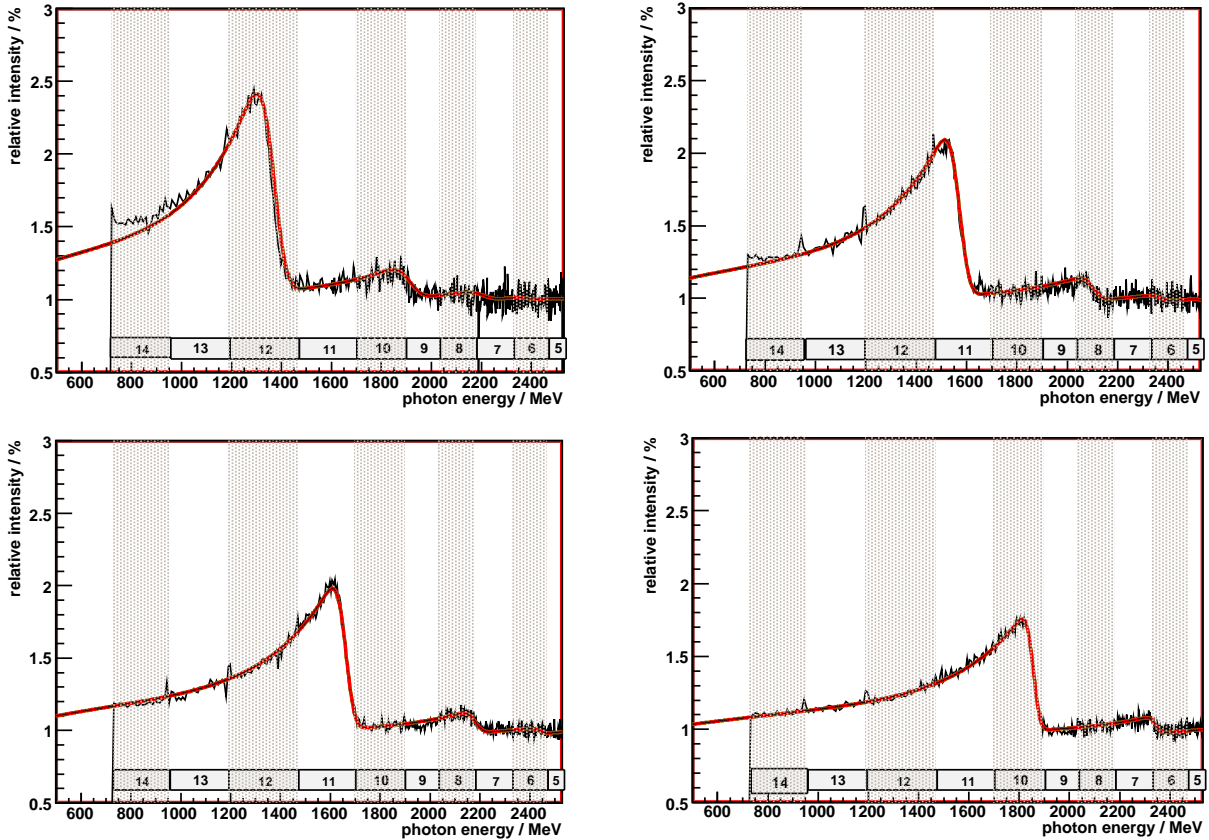}
    }
  \end{center}
\caption{The measured coherent bremsstrahlung intensity 
         normalised to an incoherent spectrum  
         in comparison to an improved version \cite{Elsner06}
         of the ANB-calculation \cite{ANB} (full curve).
         The diamond radiator was set for an intensity maximum at
         $E_\gamma=1305$\,MeV (top left), $E_\gamma=1515$\,MeV (top right),
         $E_\gamma=1610$\,MeV (bottom left), $E_\gamma=1814$\,MeV (bottom
         right). The numbered blocks indicate the ranges covered by the 14
         timing scintillators of the tagging detector.}
\label{fig:linpol_result}       
\end{figure*}

\begin{center}
  \begin{threeparttable} [h]
      \begin{tabular}{ll}
      \hline
      electron energy & 3176.1 MeV \\
      spot size $\sigma_{horizontal}$ & 1.5 mm\\
      spot size $\sigma_{vertical}$ & 1.0 mm\\
      divergence $\sigma^p_{horizontal}$ & 0.3 mrad\\
      divergence $\sigma^p_{vertical}$ & 0.08 mrad\\
      \hline
    \end{tabular}
    \caption{Electron beam properties.}
    \label{tab_beam}
  \end{threeparttable}
\end{center}
\begin{center}
  \begin{threeparttable} [h]
    \begin{tabular}[h!]{ll}
  \hline
  crystal thickness & 0.5 mm\\
  \hline
  calculated numbers of lattice vectors & 1000 \\
  \hline
  incoherent scaling factor& 1.35 \\ 
  \hline
\end{tabular}
  \caption{Radiator properties.}
  \label{tab_radiator}
\end{threeparttable}

\end{center}
\begin{center}
  \begin{threeparttable} [h]
    \begin{tabular}[t!]{llll}
      E$_\gamma$(P$^{ max.}$)/MeV & P$^{max.}$ & $\theta_h^{cry}$/mrad &
      $\theta_v^{cry}$/mrad  \\
      \hline\noalign{\smallskip}
      1305 & 0.49 & -3.16 & -56.78 \\
      \hline\noalign{\smallskip}
      1515 & 0.42 & -4.09 & -64.00  \\
      \hline\noalign{\smallskip}
      1610 & 0.39 & -4.58 & 67.00 \\
      \hline\noalign{\smallskip}
      1814 & 0.31 & -5.88 & 76.00 \\
      \hline\noalign{\smallskip}
    \end{tabular}
    \caption{Coherent peak position, maximum degree of polarisation, $P^{max.}$,
      and crystal angles for the vertical orientation of the polarisation plane.}
    \label{tab_goni_angles}
  \end{threeparttable}
\end{center}
Additionally, collimation affects the degree
of polarisation, due to the different angular distribution for the coherent
and incoherent bs. However, no effective
collimation of the photon beam was used in the experimental.\\
During the first round of CBELSA/TAPS data taking at ELSA four different crystal
settings were used, with maximum polarisation at $E_{\gamma}$=1305 MeV,
1515 MeV, 1610 MeV and 1814 MeV.
Vertical orientation of the polarisation vector was chosen, since
the vertical divergence of the ELSA electron beam is about an order of magnitude
smaller than in horizontal direction.  
Normalised electron spectra are shown in fig.\,\ref{fig:linpol_result}.
The curves represent a calculation using an improved version \cite{Elsner06} of 
the original ANB software. The description of the measured intensity spectrum
is very accurate at all settings.
Two main improvements in the ANB code were 
necessary to obtain this level of agreement between calculation and 
experimental data.
The inclusion of multiple scattering was improved 
by a more precise approximation of the angular distribution \cite{LD91}.
The original description \cite{H51} only accounts for the first order of the 
series expansion of the scattering angle in Moli\`ere theory.
This accuracy was not sufficient to describe the experimental
spectrum. As a consequence a discrepancy appears in the steep edge of the coherent peak.\\
Furthermore, the {\it incoherent} description of the ANB software needs to be
scaled for all calculations by a factor of $1.35$ (cf. table \ref{tab_radiator}).
This was traced back to an uncertainty in the parametrisation of the atomic
form factor. 
A scaling of the atomic form factor according to Cromer and Waber \cite{CW65} by a
factor of $1.1$ yields the same result, as the scaling of the
incoherent part.  Taking into account the form-factor parametrisation after Schiff
\cite{Sch51}, the difference of the form factors is also a factor of $1.1$. 
Consequently, the two alternative parametrisations provide an
uncertainty in the order of 10\% in the required momentum-transfer region. 
The relative strengths of coherent and incoherent contributions determine the
absolute value of linear polarisation. In this respect the 
re-scaling of the incoherent contributions introduces no significant
additional error.\\ 
An absolute error of $\delta P_\gamma < 0.02$ is estimated using variations of
the calculated relative intensity by $\pm5\%$. These worst-case estimate 
accounts for deviations from the shape of the spectrum 
due to combined statistical and systematical effects.\\
Assigning the appropriate photon energy to each single event in the
data analysis yields the event-weighted average polarisation in each bin of photon energy.
An independent cross check of the determination of the polarisation degree
is the measurement of the photon beam-asymmetry, $\Sigma$, 
in an energy region were it is well known from other independent measurements.
This is discussed in the next section.

\section{Beam asymmetry in {\boldmath$\pi^0$} photoproduction}
\label{sec:photoprod}
The analysis of the photon beam-asymmetry, $\Sigma$, in the reaction 
$\vec{\gamma} p \rightarrow p \pi^0$ provides a well suited
consistency check for the determination of the degree of polarisation. 
Large photon asymmetries are involved, especially at small
angles $\theta_{\pi^0}^{cm}$, and a zero-crossing at certain energies. 
Comparing our results to earlier measurements at GRAAL\footnote{{\bf GR}enoble {\bf A}nneau
  {\bf A}ccelerateur {\bf L}aser} \cite{GraalPi} gives an independent check,
in particular because at GRAAL linearly polarised photon beams are produced
by a different process, laser backscattering, with a well defined and high
degree of polarisation. 
Our results are based on the same data set and data analysis
presented in a previous publication on the beam asymmetry in $\eta$
photoproduction \cite{Elsner07}.  The experimental setup
and the main steps of the data analysis are shortly summarised in the following.

\subsection{Experimental setup and data analysis}
\label{subsec:exp_ana}
\begin{figure}[t]
  \resizebox{1.\columnwidth}{!}{%
    \includegraphics{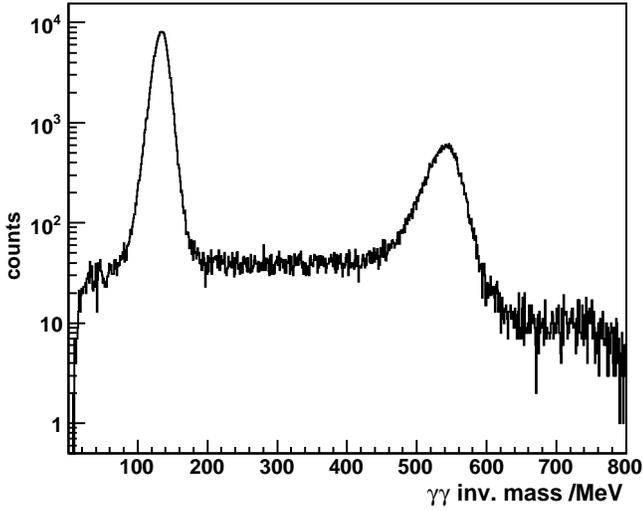} }
  \caption{Two photon invariant mass distribution 
    from the full energy and angular range
    after the standard kinematic analysis cuts (see text).
    Signal widths of $\sigma_{\pi^0} = 10$ MeV and 
    $\sigma_{\eta} = 22$ MeV are obtained.
    Note the logarithmic scale.}
  \label{fig:inv_mass}       
\end{figure}
The linearly polarised photon beam from the tagging system
(cf. Sec. \ref{sec:setup}) was incident on a $5.3$ cm long liquid hydrogen
target. The target is surrounded by a cylindrical, three layer scintillating fibre
detector, covering the polar angular range from 15 to 168 degrees,
and the \texttt{Crystal Barrel} (CB) detector \cite{CBarrel}, consisting of 1290
individual CsI(Tl) crystals covering a polar angular range of 30
--- 168 degrees.
The forward cone of $5.8$ --- 30 degree was covered by the 
\texttt{TAPS} detector, a setup of 528 BaF$_2$ modules at a
distance of $118.7$\,cm from the target.
Charged particle recognition is obtained by the hit information from the
plastic-scintillator modules in front of \texttt{TAPS} and the scintillating
fibre detector inside CB. The first level trigger was derived from the
\texttt{TAPS} detector modules, which are individually equipped with photomultiplier readout. 
Two alternative trigger conditions were used, either $\geq 2$ hits above a low threshold ($A$) or
$\geq 1$ hit above a high threshold ($B$). 
Within $\simeq 10\,\mu$s, a fast cluster recognition for the \texttt{Crystal Barrel}
provides the second level trigger ($C$).
Finally, the total trigger condition required 2 clusters identified: $[A \lor
(B \land C)]$.\\
The offline analysis is based on three detector hits, corresponding to two
photons from the pion decay, and the proton. A photon hit is usually
composed of a cluster of adjacent crystals whose energy is summed over.  
Due to the detection of the proton, the kinematics is overdetermined.
The analysis starts with all combinatorial possibilities, {\it i.e.} 3 for the 3--cluster events.
No charged particle identification for the proton was used to avoid false
azimuthal distributions due to inefficiencies of the veto detectors.
Furthermore, only the angular information of the proton candidate was used.
The energy of the proton candidate was disregarded.\\ 
Kinematic cuts, based on longitudinal and transverse momentum conservation, 
are used to extract the desired reaction.
Additionally, a cut on the missing mass was applied to the proton candidates
(m$_p\ \pm$ 150MeV).
Figure \,\ref{fig:inv_mass} shows the two-photon invariant mass distribution
\cite{Elsner07} obtained. Below the $\pi^0$ and $\eta$ peaks the overall
background is very small (note the logarithmic scale). 
After background subtraction a clean event sample was obtained from cuts of $3 \sigma$ width around the $\pi^0$ mass 
in the invariant mass spectra.\\
The cross section of pseudoscalar meson photoproduction
off a nucleon with linearly polarised beam \cite{DKT99} is given by
\begin{equation}
\frac{d\sigma}{d\Omega} = \frac{d\sigma_0}{d\Omega}\:
                          \left( 1 - P_\gamma\,\Sigma\,\cos 2\Phi \right).
\label{eq:xsec}
\end{equation}
the beam asymmetry, $\Sigma$, can be extracted from the modulation 
of the cross section over the azimuth.
In eq. \ref{eq:xsec} $\sigma_0$ denotes the polarisation independent
differential cross section,
$P_\gamma$ the degree of linear polarisation of the incident photon beam,
and $\Phi$ the azimuthal orientation of the reaction plane 
with respect to the plane of linear polarisation.
From a fit of the azimuthal event distribution
\begin{equation}
f(\Phi) = A + B\,\cos (2\Phi)
\label{eq:fit}
\end{equation}
the product of beam asymmetry and photon
polarisation, $P_\gamma \Sigma$, is given by the ratio $B/A$ in each bin of
photon energy and pion angle, $\theta^{cm}_{\pi^0}$. 
Finally, the event-weighted average polarisation, assigned as described in
Sec. \ref{sec:data}, allows the determination of $\Sigma$ in each data bin.

\subsection{Results}
\label{subsec:results}
\begin{figure*}
  \begin{center}
    \resizebox{1.\textwidth}{!}{%
      \includegraphics{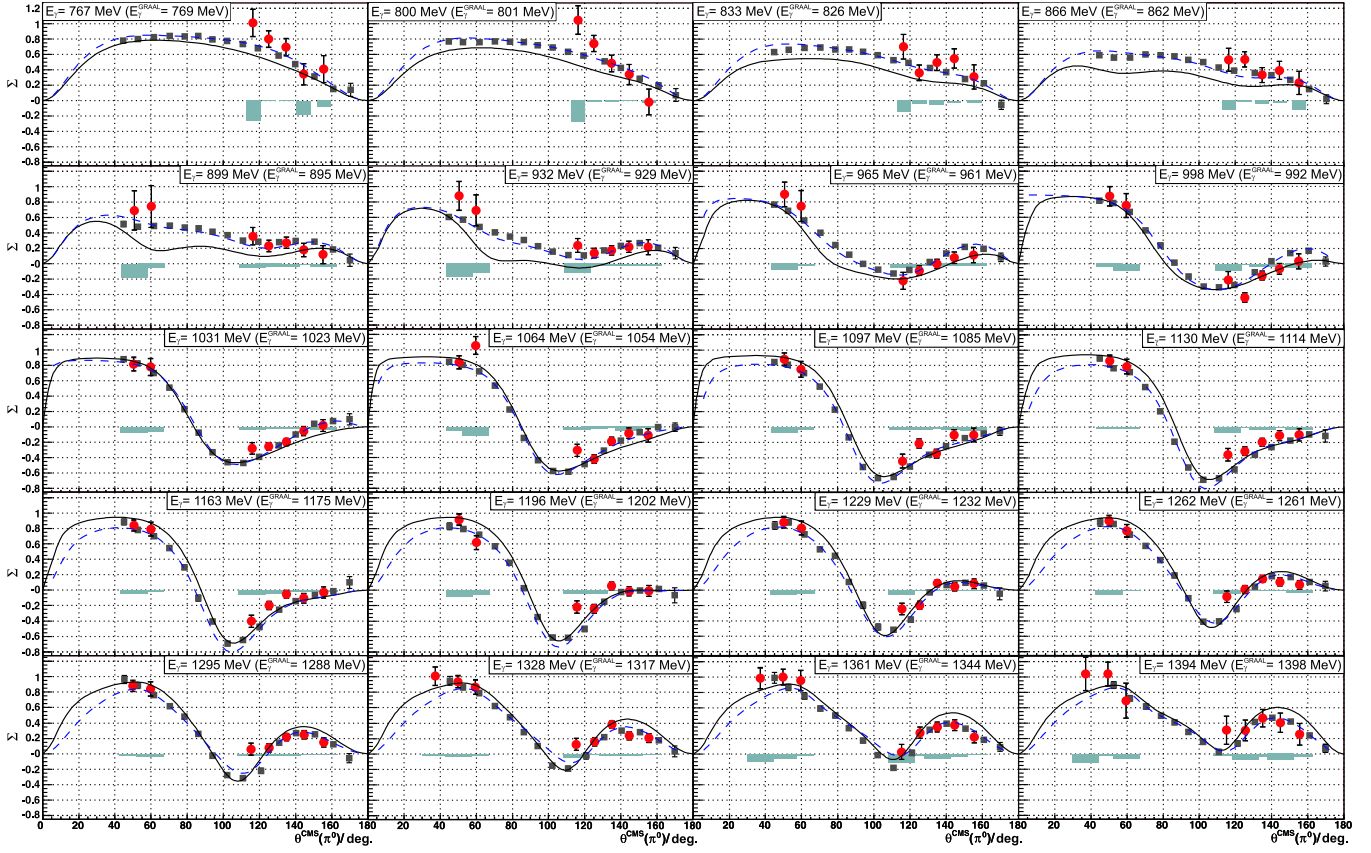}
    }
  \end{center}
  \caption{The beam asymmetry $\Sigma$ {\it versus} $\theta_{\pi^0}^{cm}$ for the reaction
    $\vec{\gamma} p \rightarrow p\pi^0$. CBELSA/TAPS data (circles) with
    statistical errors. The systematic error \cite{Elsner07} is indicated by the 
    bar chart. Our data are compared to data of the GRAAL
    collaboration \cite{GraalPi} (boxes). The curves represent calculations of \texttt{MAID} \cite{CYTD02} (full)
    and  Bonn--Gatchina partial wave analysis \texttt{BnGa} \cite{Anisovich05}
    (dashed). Data within the range $\theta_{\pi^0}^{cm}$ = 60 -- 100 degree are missing
due to the trigger condition, see text.}
  \label{fig:Sigma_Pion}       
\end{figure*}

\begin{figure}[]
  \resizebox{0.95\columnwidth}{!}{%
    \includegraphics{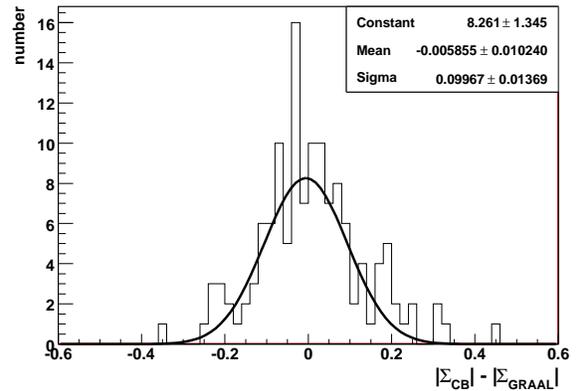} }
  \caption{The difference of the absolute values of the beam asymmetries from
    our experiment and the results from the GRAAL collaboration
    \cite{GraalPi}. Each difference of two data points corresponds to one entry in the
    histogram. In order to ensure the same kinematics of both data sets the GRAAL data are
    linearly interpolated in E$_{\gamma}$ and $\theta^{CMS}(\pi^0)$ to our
    mean values. }
  \label{fig:VsGraal}       
\end{figure}

Figure \ref{fig:Sigma_Pion} shows the results for the beam asymmetry extracted
for one crystal setting with a maximum degree of polarisation at E$_\gamma$ = 1305
MeV. Statistical errors are directly attached to the data points. The bars indicate
the estimated total systematic uncertainty. The major contribution to the
systematic error of this experiment stems from angle-dependent inefficiencies \cite{Elsner06}.
Data within the range $\theta_{\pi^0}^{cm}$ = 60 -- 100 degree are missing
due to the trigger condition $B$, which was used to select events with higher photon multiplicity
than $\pi^0\rightarrow 2\gamma$. 
In fig. \ref{fig:Sigma_Pion} our data are compared to data of the GRAAL
collaboration \cite{GraalPi}. 
Both data sets show a very good agreement, despite
small fluctuations around  $\theta_{\pi^0}^{cm}$ = 115 degree. 
The kinematics of these data points are correlated to a very low proton energy,
probably the protons got stuck in the $\phi$-unsymmetrical holding structure of the inner
detector. This $\phi$-dependence of detection efficiency affects directly the experimental asymmetry.\\
Figure \ref{fig:Sigma_Pion} also shows the good agreement between both
data sets and two standard calculations, the Mainz isobar model \texttt{MAID} \cite{CYTD02}  
and the Bonn--Gatchina partial wave analysis \texttt{BnGa} \cite{Anisovich05}.\\
An explicit deviation of both data sets is shown in fig. \ref{fig:VsGraal}. 
The difference of the absolute values of the beam asymmetries ($|\Sigma_{CB}| -
|\Sigma_{GRAAL}|$) is plotted in one histogram. This representation is more
sensitive to an incorrect measurement of the photon polarisation than the
difference of the signed values. 
In fig. \ref{fig:VsGraal} the mean value of a gauss distribution
is compatible with zero. Also the width (sigma) corresponds to
our mean statistical error. Consequently this cross check shows no indications for an
additional contribution to the systematic error in the determination of the
degree of polarisation. Hence we conclude that the absolute determination of
the degree of linear polarisation is under control on the level of the
estimated errors given in sect. 5.\\

\section{Summary}
\label{subsec:summary}
We have presented the method to produce a linearly polarised photon
beam at ELSA by means of coherent brems-strahlung off a diamond crystal. 
Within the photon energy range $E_\gamma = 800 ... 1400$\,MeV we achieve
polarisation degrees up to 49\,\%. At higher energies the polarisation reduces
to {\it e.g.} P$_{\gamma}$ $\approx$ 30\% at E$_{\gamma}$ = 1800 MeV. 
The precise orientation of the diamond crystal {\it versus} the incoming
electron beam is essential. It is realised by a 5-axis goniometer. The
alignment is based on the {\it Stonehenge Technique}.
Both, the relative intensity spectrum and the polarisation degree,
have been calculated with an improved version of the ANB software.\\
An independent  consistency check is provided by the measurement of the photon
beam asymmetry, $\Sigma$, in the reaction $\vec{\gamma} p \rightarrow p \pi^0$. 
The combined setup of the \texttt{Crystal Barrel}
and \texttt{TAPS} detectors enabled a high-resolution detection 
of multiple photons, important for the clean detection of the 
$2\gamma$ decays of the pion.
The obtained photon asymmetries are compared with a previous
measurement by the GRAAL collaboration.
A very good overall consistency of the data sets is obtained.
No deviations were visible beyond the error of $\delta P_\gamma < 0.02$ given
above. The production of linearly polarised photons via coherent bremsstrahlung and
the presented method of determination of the degree of polarisation is now
routinely used as a standard technique at ELSA.\\

\begin{acknowledgement}
We are happy to acknowledge the continuous efforts of the accelerator
crew and operators to provide stable beam conditions.
K. Livingston from Glasgow university deserves a big share
of credit for his invaluable help in setting up the Stonehenge technique
for the crystal alignment.
This work was financially supported by the federal state of 
{\em North Rhine-Westphalia} and the
{\em Deutsche Forschungsgemeinschaft} within the SFB/TR-16.
The Basel group acknowledges support from the
{\em Schweizerischer Nationalfonds}, 
the KVI group from the {\em Stichting voor Fundamenteel Onderzoek der 
Materie} (FOM) and the {\em Nederlandse Organisatie voor Wetenschappelijk 
Onderzoek} (NWO).
\end{acknowledgement}

%

\end{document}